\documentclass[twocolumn,aps,prl,showpacs,amsmath,amssymb]{revtex4}
\usepackage{graphics}
\usepackage{epsf,graphicx,pstcol,epsfig,psfrag}
\begin{document}

\title{Quantum criticality in Ce$_2$PdIn$_8$: thermoelectric study}
\author{Marcin Matusiak, Daniel Gnida, and Dariusz Kaczorowski}
\affiliation{Institute of Low Temperature and Structure
Research, Polish Academy of Sciences, P.O. Box 1410, 50-950 Wroc{\l}aw, Poland}
\date{\today}

%%%%%%%%%%%%%%%%%%%%%%%%%%%%%%%%%%%%%%%
\begin{abstract}
We report the Nernst effect ($\nu$) and thermoelectric power ($S$) data for the Ce$_2$PdIn$_8$ heavy-fermion compound. Both $S$ and $\nu$ behave anomalously at low temperatures: the thermopower shows a Kondo-like maximum at $T$ = 37 K, while the Nernst coefficient becomes greatly enhanced and field dependent below $T \approx$ 30 K. In the zero-$T$ limit $S/T$ and $\nu/T$ diverge logarithmically, what is related to occurrence of the quantum critical point (QCP). Presented results suggest that the antiferromagnetic spin-density-wave scenario may be applicable to QCP in Ce$_2$PdIn$_8$.
\end{abstract}


\maketitle
%%%%%%%%%%%%%%%%%%%%%%%%%%%%%%%%%%%%%%%%%%%%%%%%%%
%\section{Introduction}
%%%%%%%%%%%%%%%%%%%%%%%%%%%%%%%%%%%%%%%%%%%%%%%%%%%
The physical properties of matter in the vicinity of quantum critical point (QCP) has been intensively studied for the last two decades. A relation between QCP and unconventional superconductivity, as well as other unusual properties of the electronic system near QCP have attracted much attention, although this complex topic is far from being well understood. Measurements of the thermoelectric effects, which are particularly sensitive to distribution of the density of the electronic states and variation of the relaxation time around the Fermi surface, turn out to be an effective method to detect departures from the Landau description of metals \cite{Behnia,Bel,Cyr,OnoseIzawa,Bel2}. However, an interpretation of these anomalies is not straightforward, since there exist a wide variety of overlapping to some extent models that attempt to explain the unusual thermoelectric behavior. For instance, one can mention here the models of d-wave- (dDW) , spin- (SDW), charge- (CDW), and unconventional- (UDW) density-waves \cite{Oganesyan,DW}, magnetic-field-induced chiral order \cite{Kotetes}, superconducting vortices (or vortex-like excitations) \cite{Xu}, superconducting fluctuations \cite{Ussishkin}, current vertex correction \cite{Onari}. Even though, studies on temperature and/or magnetic field variations of the Nernst coefficient and the thermoelectric power have been recognized as very helpful tools in distinguishing between possible types of quantum criticality \cite{Paul,Hartmann,Kim}.

In this Letter we investigate the transport properties of the Ce$_2$PdIn$_8$ heavy-fermion (HF) compound that exhibits unconventional superconductivity below $T_c=0.7$ K \cite{prl}. We focus on the thermoelectric phenomena which appear to be highly anomalous in the low temperature region. Both the Nernst effect and the thermoelectric power show below $T \approx$ 7 K features that can be attributed to presence of the quantum critical point.

%section{Experimental}
High-quality polycrystalline sample of Ce$_2$PdIn$_8$ was prepared and characterized as described in Ref. \cite{Kaczorowski}. The resistivity was measured using the four-probe technique with 25 $\mu$m gold wires attached to the bar-shaped sample with two component silver epoxy (EPO-TEK H20E). For the Hall coefficient measurement, the sample was mounted on a rotatable probe and continuously turned by 180 degree (face down and up) in a magnetic field ($B$) of 12.5 T to effectively reverse the field anti-symmetrical signal. During the thermoelectric power and Nernst coefficient measurements, the sample was clamped between two phosphor bronze blocks, which had two Cernox thermometers and resistive heaters attached to them. The temperature runs were performed in magnetic fields from -13 to +13 T in order to extract the field voltage components that were odd and even in $B$.

%%%%%%%%%%%%%%%%%%%%%%%%%%%%%%%%%%%%%%%%%%%%%%%%%%%%%%
%\section{Results}
%%%%%%%%%%%%%%%%%%%%%%%%%%%%%%%%%%%%%%%%%%%%%%%%%%%%%

Ce$_2$PdIn$_8$ was reported to exhibit unusual transport properties at low temperatures. Both the thermal conductivity \cite{Dong} and the electrical resisitivity ($\rho$) \cite{Dong,Tran} indicate the presence of an antiferromagnetic (AF) quantum critical point induced by magnetic field of $B \approx 2.4$ T. For this field the resistivity changes below temperature $T \approx 7$ K as $\rho(T) = \rho_0 + AT$ down to $T$ = 50 mK, while for fields lower than 2.4 T the linear $\rho(T)$ trend is ceased by the superconducting transition ($T_c=0.7$ K at $B$ = 0 T). Since the low temperature linear $\rho(T)$ dependence is considered to be a hallmark of non-Fermi-liquid (NFL) state \cite{Stewart}, one can anticipate deviations from regular metallic behavior in other physical properties as well.
The low-$T$ part of the presented in Fig. 1 ($a$) temperature dependence of the Hall coefficient ($R_H$) is an example of such a deviation.
\begin{figure}
\label{fig1}
 \epsfxsize=8.5 cm \epsfbox{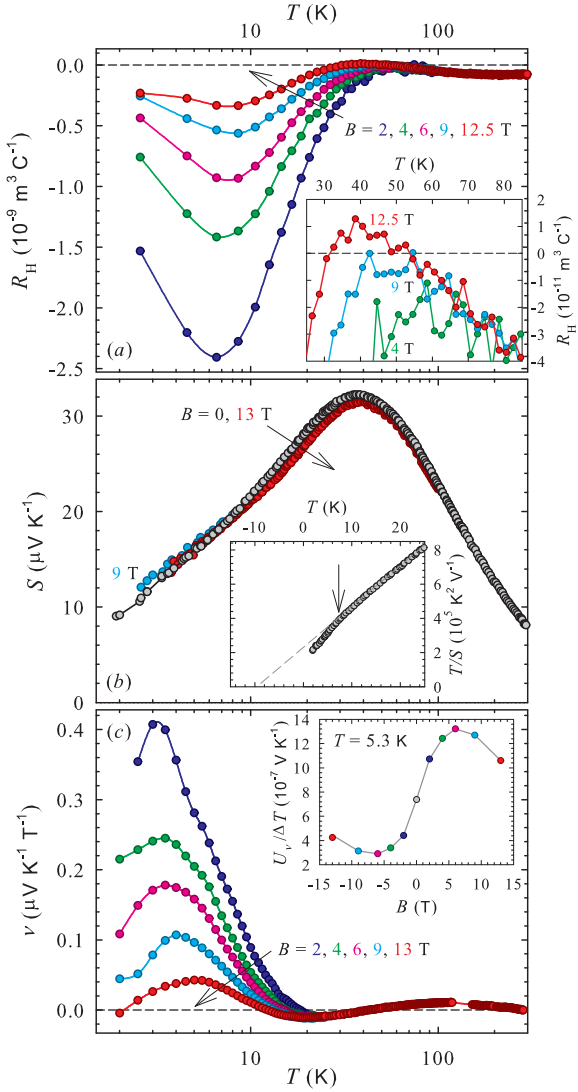}
 \caption{(Color online) Temperature dependences of the transport coefficients for various magnetic fields in semi-log plots. Panel $(a)$ presents the Hall coefficient, and inset in this panel shows $R_H(T, B$=4T), $R_H(T, B$=9T) and $R_H(T, B$=12.5T) in the temperature region, where the latter changes sign for positive. Panel $(b)$ presents the thermoelectric power, and inset in this panel shows the $T/S$ versus $T$ plot used to estimate $T_{coh}$. The arrow indicates temperature $T \approx 7$ K, below which data departures from the fit. Panel $(c)$ presents the Nernst coefficient, and inset in this panel shows an example of the non-linear field dependence of the Nernst signal, which is the transverse voltage divided by longitudinal temperature gradient.}
 \end{figure}
Below $T \approx 50$ K, $R_H$ becomes field dependent and its magnitude is significantly enhanced. Very similar behavior of $R_H(T,B)$ in CeCoIn$_5$ and CeRhIn$_5$ was attributed to antiferromagnetic spin fluctuations causing highly anisotropic scattering time \cite{Nakajima}. For the highest applied magnetic field ($B=$ 12.5 T), $R_H$ changes sign for positive for $30 \lesssim T \lesssim 50$ K. This can be a result of skew-scattering or/and multiband compensation. Above $T\approx$ 50 K, the Hall coefficient is no longer field dependent and $R_H$ decreases slowly with increasing temperature to reach a value of -8 x $10^{-11}$ m$^3$/C at room temperature.

Figure 1 ($b$) presents the temperature dependences of the Seebeck coefficient ($S$) in Ce$_2$PdIn$_8$, measured in zero and finite magnetic field. The observed enhanced value of $S$ and overall shape of the $S(T)$ curve are typical of Kondo scattering, in line with the previous characterization of the compound as a Kondo system \cite{Kaczorowski}. In a single-impurity case, the position ($T_{max}$) of the broad peak in $S(T)$ is often identified as an estimate for the Kondo temperature ($T_K$) \cite{Hewson}. For a Kondo lattice it was suggested that above the coherence temperature ($T_{coh}$) magnetic ions can be treated as independent scattering centers \cite{Zlatic}, and one can expect occurrence of a peak in $S(T)$ in the vicinity of $T_{coh}$ instead of $T_K$ \cite{Dilley}. For Ce$_2$PdIn$_8$, the thermopower peaks at $T_{max} \approx 37$ K, which is fairly close to the value previously indicated on the basis of the electrical resistivity data $T_{coh} = 29$ K \cite{Kaczorowski}. Both values are nearly insensitive to the magnetic field. Namely, $T_{max}$ increases to about 39 K at $B$ = 13 T, while $T_{coh}$ decreases in the magnetic field of 12 T to approximately 27 K \cite{Tran}. The opposite direction of the $T_{coh}(B)$ and $T_{max}(B)$ changes are likely due to minor field sensitive additions (e.g. related to AF fluctuations or phonon-drag) to the dominating Kondo thermopower that is almost field independent. The same explanation can be applied to the small $S(B)$ dependence at low temperatures, where, for instance, a difference between $S(T,B$ = 9 T) and $S(T,B$ = 0 T) reaches only about 10\% at $T$ = 3 K.
 
Another method to estimate $T_{coh}$ can be adapted from the empirical observation that for $T \lesssim T_K$ the thermopower related to the single-ion Kondo effect ($S_K$) varies like $S_K = AT / (T + 0.35 T_K)$ \cite{Cooper}. By replacing in this formula $T_K$ by $T_{coh}$, one obtains [cf. the extrapolation shown in the inset to Fig. 1(b)] $T_{coh} \approx 28$ K that matches the value derived from $\rho(T)$.  

The field and temperature dependences of the Nernst coefficient ($\nu$) shown in Fig. 1($c$) are similar to the $R_H(T,B)$ variations, except for the opposite sign. We use the Nernst effect sign convention, where vortex flow in a superconductor gives positive $\nu$ \cite{Behnia}. Above $T_{coh}$, the Nernst coefficient is field independent and very small. At room temperature $\nu$ is of the order of 1 nV/KT. Upon cooling the Nernst coefficient increases to about 10 nV/KT at $T \approx$ 110 K, then it starts to decrease and crosses zero at $T$ = 43 K. Given the multiband electronic structure of other Ce$_n$TIn$_{3n+2}$ compounds ($n$ = 1, 2 and T = Co, Rh, Ir) \cite{Haga}, this small magnitude of $\nu$ in the wide temperature range may be surprising. In the presence of positive (denoted below with index $p$) and negative (index $n$) charge carriers one can expect to observe an enhancement of the Nernst coefficient due to the ambipolar flow of quasiparticles \cite{Bel,Matusiak}. If the ambipolarity played a role in the thermoelectric transport, then $\nu$ should be maximal the region, where the electronic transport is completely compensated \cite{Delves}. In contrast, the Nernst coefficient in Ce$_2$PdIn$_8$ crosses zero  ($\nu$ = 0 nV/KT in  $B$ = 13 T at $T$ = 43 K, Fig. 1 ($c$)) in the temperature region, where the Hall coefficient is very small ($R_H$ = 0 m$^3$/C in $B$ = 12.5 T at $T \approx$ 30 and 50 K , inset in Fig. 1 ($a$)). In two band model the total Nernst coefficient reads \cite{Sondheimer}:
\begin{equation}
\nu = \frac{\nu_p \sigma_p + \nu_n \sigma_n}{\sigma_p + \sigma_n} + \frac{\sigma_p \sigma_n (S_p - S_n) (R_{Hp} \sigma_p - R_{Hn} \sigma_n)}{(\sigma_p + \sigma_n)^2},
\end{equation}
where $\sigma$ is the electrical conductivity. This indicates a dominating role of the majority charge carriers in electrical transport, since the second term in Eqn. 1 is small if $\sigma_p \ll \gg \sigma_n$.
A dramatic change of the Nernst signal occurs below $T\approx$ 30 K, where the value of $\nu$ significantly rises and becomes field dependent (an example of nonlinear $B$-dependence of the Nernst signal at $T$ = 5.3 K is given in the inset in Fig. 1 ($c$)). Intriguingly, the onset of the field dependent behavior occurs in the Hall effect at the temperature about 20 K higher than that in the Nernst effect. Because the latter is considered to be exceptionally sensitive to an emerging order in the electronic system \cite{Cyr}, this would suggest rather local character of the magnetic features. Very similar low temperature $\nu(T,B)$ dependence was reported in CeCoIn$_5$ \cite{Bel} and URu$_2$Si$_2$ \cite{Bel2}. In URu$_2$Si$_2$ the thermopower and the Hall coefficient have the opposite sign to those presented here, whereas in CeCoIn$_5$ all three coefficients have the same signs as in Ce$_2$PdIn$_8$. The two-band model was judged to be unable to explain this kind of behavior \cite{Nakajima}, but even in a one band picture there are difficulties in clear interpretation of the Nernst coefficient sign. In fact, the sign of $\nu$ is determined together by type (electron- or hole-like) of charge carriers, sign of the energy derivative of the relaxation time ($\tau$), and also by the energy dependence of the effective mass \cite{Behnia}.

The observed variation in the magnitude of $\nu$ in the 7 K $\lesssim T \lesssim 30$ K region can be attributed to strongly anisotropic relaxation time $\tau$. It was shown that in case of a non-parabolic conductivity band the Nernst coefficient can be enhanced  even if $\tau$ is only momentum dependent \cite{Clayhold}: $\nu = [\langle S \tan \theta \rangle - \langle S \rangle \langle \tan \theta \rangle]/B$. Here $\tan \theta$ is the Hall angle, and $\langle \rangle$ refer to Fermi surface averages weighted by the electrical conductivity. Because the normal-state Nernst coefficient is composed of two terms $\nu = (\rho \alpha_{xy} - S \tan \theta)/B$, where $\alpha_{xy}$ is the off-diagonal element of the Peltier tensor \cite{Wang}, it is necessary to check, whether the observed strong temperature and field dependence of $\nu$ is not a product of the variation of $\rho$, $R_H$ and $S$. Fig. 2 presents $(S \tan \theta)/B$ calculated for $B$ = 2 and 13 T \cite{explanation}, and in both fields the temperature variation of $S \tan \theta$ appears to be much weaker than $\alpha_{xy}(T)$.
\begin{figure}
\label{fig2}
 \epsfxsize=8 cm \epsfbox{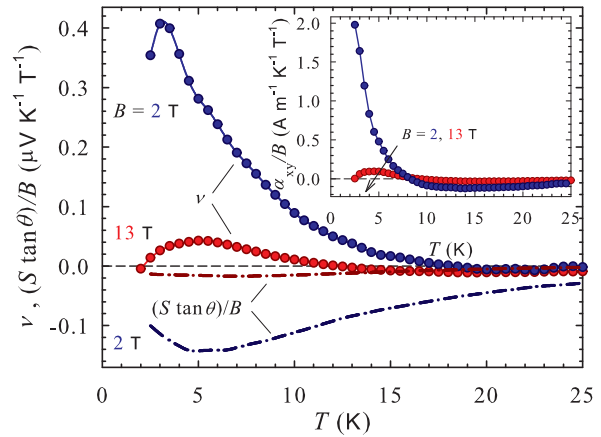}
 \caption{(Color online) Temperature dependences of the Nernst coefficient (points) and $(S \tan \theta)/B$ term (dash-dotted lines) in Ce$_2$PdIn$_8$ measured at $B$ = 2, and 13 T. Inset shows off-diagonal element of the Peltier tensor for the same magnetic fields and temperature range.}
 \end{figure}
Indeed, $ \alpha_{xy}(T)/B$ calculated for B = 2 and 13 T (see the inset in Fig. 2) appears to be at low temperatures a strong function of $T$ and $B$. For both, low and high, fields $\alpha_{xy}(T)/B$ slowly decreases upon cooling, then it starts to rise at $T \approx$ 15 K, and $\alpha_{xy}$ changes sign from negative to positive at $T \approx$ 8.5 K. In a field of 13 T this variation is reversed at $T \approx$ 4.5 K, whereas at $B$ = 2 T $\alpha_{xy}(T)$ maintains its divergent trend when nearing zero temperature. Such a behavior of $\alpha_{xy}$ implies dramatic changes in the Hall conductivity ($\sigma_{xy}$), since the off-diagonal element of the Peltier tensor is the energy derivative of $\sigma_{xy}$ at the chemical potential ($\mu$): $\alpha_{xy} = -\frac{\pi^2 k_B^2}{3e} (\frac{\partial \sigma_{xy}}{\partial \varepsilon})_{\mu}$ \cite{Oganesyan}, where $k_B$ is the Boltzmann constant, and $e$ is the elementary charge. Because the Hall conductivity is a quantity sensitive to changes in the topology of the Fermi surface \cite{Kee}, this finding hints at a distinct reconstruction of the Fermi surface in Ce$_2$PdIn$_8$ in the low temperature and relatively weak field limit. The relation between the Nernst coefficient, Hall mobility ($\mu_H$) and Fermi energy ($\varepsilon_F$) \cite{Behnia},

 $\frac{\nu}{T} = -\frac{\pi^2 k_B^2}{3e}\frac{\mu_H}{\varepsilon_F}$

,  was shown to hold in the zero-temperature limit in various metals including strongly correlated ones. This relation is approximately valid also in multi-band systems, despite the fact that the Fermi energy and the mobility can be different in different bands. In the lower inset to Fig. 3 there is shown the low temperature variation of the Fermi energy estimated for Ce$_2$PdIn$_8$ with the aid of the above formula. Remarkably, in $B = 2$ T $\varepsilon_F$ turns out to be extremely small (of the order of 1 meV) and it gradually decreases toward zero for $T \rightarrow 0$. In other words, near the anticipated quantum critical point the characteristic energy scale becomes very small, and may collapse in the zero-$T$ limit.
\begin{figure}
\label{fig3}
 \epsfxsize=8 cm \epsfbox{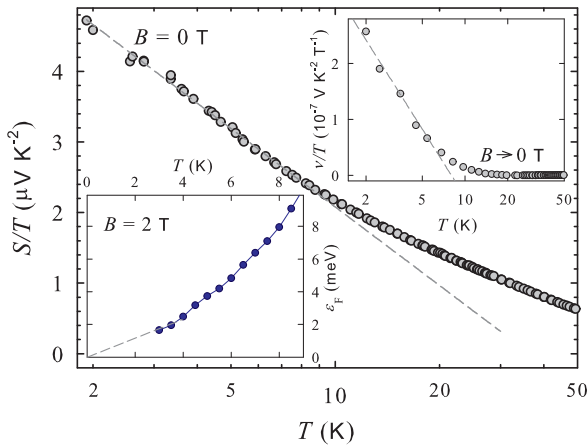}
 \caption{(Color online) Thermopower in Ce$_2$PdIn$_8$ divided by temperature in a semi-log plot. The dashed line is the low temperature logarithmic fit $S/T \propto \ln(1/T)$. Upper inset presents the similar $\nu / T$ dependence, where the Nernst coefficient was estimated from the slope of  $\nu(B)$ in the zero-field limit. The dashed line is the logarithmic fit. Bottom inset shows the Fermi energy in Ce$_2$PdIn$_8$ as a function of temperature for $B$ = 2 T. The dashed line is a linear extrapolation of $\varepsilon_F(T)$ down to $T$ = 0 K.}
 \end{figure}

The behavior of $\varepsilon_F(T)$ must influence the electronic transport properties, and for instance, a divergent dependence of $S/T$ is expected to appear in the vicinity of the QCP \cite{Paul,Hartmann}. Such a behavior is shown in the inset in Fig. 1 ($b$), where $T/S$ below $T \approx$ 7 K deviates from a linear dependence. At low temperatures, the thermopower of ordinary metals as well as Kondo systems (for $T \lesssim 0.1 - 0.15 T_{coh}$ \cite{Cooper}) should be $S \propto T$, thus the value of $T/S$ saturates in zero-$T$ limit. In contrast, $T/S$ in Ce$_2$PdIn$_8$ starts to incline even faster below $T \approx$ 7 K (see the inset to Fig. 1b). As may be inferred from Fig. 3, in this temperature region one observes a logarithmic relation $S/T \propto \ln(1/T)$ that may be considered as a fingerprint of underlying quantum criticality \cite{Kim}. Moreover, the same kind of functional dependence seems to be also applicable to $\nu/T$ presented in the upper inset in Fig. 3 (here the Nernst coefficient was estimated in the extrapolation to the $B$ = 0 T limit). A logarithmic increase of $S/T$ upon cooling together with linear $\rho(T)$ is expected to be observed in the case of the itinerant two-dimensional spin-density-wave (SDW) quantum criticality \cite{Paul}. A rival explanation within the framework of the local Kondo-breakdown theory predicts the same type of variations for three-dimensional systems \cite{Kim}. However, the layered crystal structure of Ce$_2$PdIn$_8$, and similarities to other members of the Ce$_n$TIn$_{3n+2}$ family that exhibit quasi 2D properties, prompts us to conclude that Ce$_2$PdIn$_8$ belongs to the same class of field-induced AF SDW quantum critical systems as CeCoIn$_5$. Additionaly, the emerging at $T_c=0.7$ K superconductivity supports our conclusion, since critical fluctuations of a spin-density wave QCP were suggested to promote unconventional superconductivity \cite{Si}.

%%%%%%%%%%%%%%%%%%%%%%%%%%%%%%%%%%%%%%%%%%%%%%%%%%
%\section{Summary}
%%%%%%%%%%%%%%%%%%%%%%%%%%%%%%%%%%%%%%%%%%%%%%%%%%%
In summary, the unusual thermoelectric properties of Ce$_2$PdIn$_8$ manifest its anomalous metallic state. The temperature dependencies of all the measured quantities may be roughly divided into three regions, where different mechanisms play major role in determining characteristics of the transport phenomena: (i) high temperature region ($T \gtrsim$ 50 K), where Ce$_2$PdIn$_8$ behaves like a Kondo metal; (ii) medium temperature region (7 K $ \lesssim T \lesssim $ 50 K), where strong field and temperature dependences of $R_H$ and $\nu$ are caused by anisotropic scattering time due to antiferromagnetic fluctuations; (iii) low temperature region ($T \lesssim $ 7 K), where transport properties of Ce$_2$PdIn$_8$ are determined by the underlying quantum critical point. The low temperature logarithmic divergence of $S/T$ together with linear $\rho(T)$ and the expected quasi-2D behavior in Ce$_2$PdIn$_8$ suggests that QCP in this compound belongs to the 2D SDW AF class. The low-$T$ divergent behavior of $\nu / T$ appears to be logarithmic as well, and the estimated $\varepsilon_F(T)$ extrapolates to zero at $T$ = 0 K. Remarkably, at high temperatures there is no detectable ambipolar enhancement of the Nernst signal, which suggests that Ce$_2$PdIn$_8$ is effectively one-band metal. This feature can be helpful in testing theoretical models.

%%%%%%%%%%%%%%%%%%%%%%%%%%%%%%%%%%%%%%%%%%%%%%%%%%%%%%%%%%%
\section*{Acknowledgments}
The authors are grateful to V. Zlati\'{c} for valuable discussion. This work was supported by a grant No N N202 130739 of the Polish Ministry of Science and Higher Education.
%%%%%%%%%%%%%%%%%%%%%%%%%%%%%%%%%%%%%%%%%%k%%%%%%%%%%%%%%%%%%%%%%%%


\begin{thebibliography}{}
\bibitem{explanation} For experimental reasons, $\nu$ was measured at the maximal field  $B_{max}^{\nu}$  = 13 T, $R_H$ at  $B_{max}^{Rh}$ = 12.5 T, whereas $\rho$ at $B_{max}^{\rho}$ = 12 T. Here, for the qualitative analysis we assume that $B_{max}^{\nu} \approx B_{max}^{Rh} \approx B_{max}^{\rho}$ = 13 T.
\end{thebibliography}
\end{document}